\documentclass[journal,letterpaper]{myclass}
\usepackage{cite}
\usepackage{graphicx}

\newcommand{\Fig}[1]{Fig.\,\ref{#1}}
\newcommand{\FigIns}[2]{
   \begin{figure}
   \centering
   \includegraphics[width=3.25in]{#1}
   \caption{#2}
   \label{#1}
   \end{figure}
   }

\newcommand{\CClass}[1]{\emph{#1}}
\newcommand{\CMethod}[1]{\emph{::#1}}

\newcommand{\PaxID}{\CClass{PaxId}}
\newcommand{\PaxName}{\CClass{PaxName}}

\newcommand{\PaxFV}{\CClass{PaxFourVector}}
\newcommand{\PaxVX}{\CClass{PaxVertex}}
\newcommand{\PaxCO}{\CClass{PaxCollision}}

\newcommand{\PaxLV}{\CClass{PaxLorentzVector}}
\newcommand{\PaxTV}{\CClass{PaxThreeVector}}

\newcommand{\PaxEI}{\CClass{PaxEventInterpret}}
\newcommand{\PaxUR}{\CClass{PaxUserRecord}}

\newcommand{\PaxMap}{\CClass{PaxMap}}
\newcommand{\PaxMapKI}[2]{\CClass{PaxMap$<$#1, #2$>$}}

\newcommand{\PaxMultiMapKI}[2]{\CClass{PaxMultiMap$<$#1, #2$>$}}

\newcommand{\PaxIterator}{\CClass{PaxIterator}}
\newcommand{\PaxMapIterator}{\CClass{PaxMapIterator}}
\newcommand{\PaxMultiMapIterator}{\CClass{PaxMultiMapIterator}}

\newcommand{\PaxRelationManager}{\CClass{PaxRelationManager}}

\newcommand{\PaxEC}{\CClass{PaxExperimentClass}}

\newcommand{\PaxFVM}{\CClass{PaxFourVectorMap}}
\newcommand{\PaxVXM}{\CClass{PaxVertexMap}}
\newcommand{\PaxCOM}{\CClass{PaxCollisionMap}}

\newcommand{\decaysto}{\rightarrow}
\newcommand{\ttH}{t\bar tH}
\newcommand{\Hbb}{H\decaysto b\bar b}

\begin{document}

\title{The PAX Toolkit and its Applications \\ at Tevatron and LHC}
\author{Steffen~Kappler,~Martin~Erdmann,~Ulrich~Felzmann,~Dominic~Hirschb\"uhl,\\ Matthias~Kirsch,~G\"unter~Quast,~Alexander~Schmidt~and~Joanna~Weng
\thanks{Manuscript submitted to IEEE Trans. Nucl. Sci., November 15, 2004, revised July 22, 2005.}
\thanks{S.~Kappler, M.~Erdmann and M.~Kirsch are with III.~Physikalisches Institut~A, RWTH Aachen, Germany.}
\thanks{U.~Felzmann, D.~Hirschb\"uhl, G.~Quast, A.~Schmidt and J.~Weng are with Institut f\"ur Experimentelle Kernphysik, Universit\"at Karlsruhe (TH), Germany.}
\thanks{Contact: steffen.kappler@cern.ch}}


\maketitle

\begin{abstract}
At the CHEP03 conference we launched the Physics Analysis eXpert
(PAX), a C++ toolkit released for the use in advanced high energy
physics (HEP) analyses. This toolkit allows to define a level of
abstraction beyond detector reconstruction by providing a general,
persistent container model for HEP events. Physics objects such as 
particles, vertices and collisions can easily be stored, accessed
and manipulated. Bookkeeping of relations between these objects
(like decay trees, vertex and collision separation, etc.) including 
deep copies is fully provided by the relation management. Event 
container and associated objects represent a uniform interface for 
algorithms and facilitate the parallel development and evaluation 
of different physics interpretations of individual events. 
So-called analysis factories, which actively identify and distinguish 
different physics processes and study systematic uncertainties, can 
easily be realized with the PAX toolkit.

PAX is officially released to experiments at Tevatron and LHC. 
Being explored by a growing user community, it is applied in a number of 
complex physics analyses, two of which are presented here. We report 
the successful application in studies of $t\bar t$ production at the 
Tevatron and Higgs searches in the channel $t \bar tH$ at the LHC and give 
a short outlook on further developments.
\end{abstract}

\section{Introduction}
\PARstart{P}{hysics} analyses at modern collider experiments enter a
new dimension of event complexity. At the LHC, for instance, physics
events will consist of the final state products of the
$\mathrm{O}(20)$ collisions taking place during each readout cycle.
In addition, a number of physics questions is studied
in channels with complex event topologies and configuration
ambiguities occurring during event analysis.

\FigIns{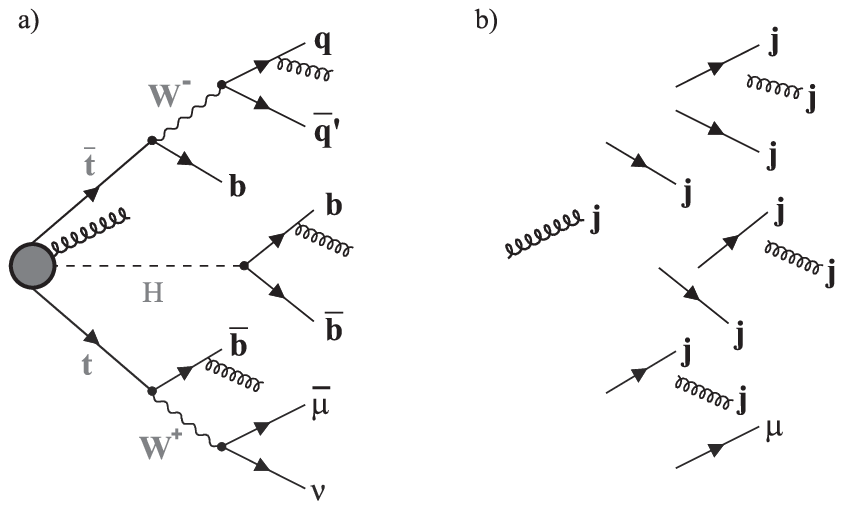}{a) Associated Higgs production in the channel $\ttH$ with $\Hbb$ and $t\bar t \decaysto WW\,b\bar b \decaysto qq'\, \mu \bar \nu_\mu\, b\bar b$. 
                     ~~b) The visible reconstructed partons of this channel.}

One item in the long list of examples is a channel of $t$-quark 
associated Higgs production, $\ttH$ with $\Hbb$ (see \Fig{ttHFeyn.eps}.a).
The event topology of four $b$-jets, two light-quark-jets,
an isolated muon, missing energy and possible additional jets from initial state radiation (ISR)
and final state radiation (FSR) imposes highest demands on detectors and reconstruction algorithms.
In addition, non-trivial ambiguities must be resolved during event analysis.
Even if all final state products could be reconstructed perfectly 
(as illustrated in \Fig{ttHFeyn.eps}.b) 
and no ISR or FSR effects occured, 
at least 24 different configurations would be possible.
Finite jet resolutions, limited efficiency and purity of
the $b$-tagging as well as the presence of additional jets 
complicate ambiguity resolution and signal identification. 

This task can be approached with a likelihood method based on 
characteristical event variables, where each possible event 
configuration is developed individually and rated with the 
likelihood function; the most probable of all interpretations 
finally is selected.

Such an approach can be implemented by object-oriented coding 
and suggests the use of a class collection, that provides 
event containers for the reconstructed objects 
(muons, jets, missing energy, vertices, collisions, etc.)
and handles relations between the individual objects 
(as, for instance, vertex relations for particle decays). 
Due to the large number of ambiguities occurring during the 
reconstruction of $\ttH$ events,
these classes are required to offer 
automated copy functionality for containers, 
objects and corresponding relations.

The application of a \emph{generalized event container} comes 
with a number of desirable side-effects.
If used to define an abstraction interface between 
the output of event generator, simulation or reconstruction 
software and the physics analysis code, the latter is 
protected from changes in the underlying software packages
to a large extent.
This reduces code maintainance and increases code lucidity.
In addition, unnecessary duplication of the analysis code can 
be avoided: so can the influence of detector effects (studied by 
direct comparison of the results on generator, simulation 
and on real data level) be investigated ad hoc, i.e.\ with the 
same analysis source code. 

Analysis factories, in which a number of analyses are executed 
at the same runtime, identifying and distinguishing different 
physics processes or studying systematic uncertainties, can easily 
be realized when using common physics objects and a common event 
container model in each of the analyses.

Analysis environments based on a well-defined, generalized event 
container also provide a basis for efficient team work. 
Collaboration in (and supervision of) groups of students is 
facilitated, and knowledge transfer between subsequent generations 
of analysts as well as between different experiments is fostered.

In this article, we present the Physics Analysis eXpert (PAX), 
a C++ toolkit for particle physics analysis that provides such a 
generalized event container together with various built-on 
functionalities. 

\section{The PAX class structure}

The PAX kernel, introduced in the year 2002 \cite{PAX02} and released
at the CHEP03 conference in 2003 \cite{PAX03}, is currently available as
2.00 version. 
For the convenience of connecting to existing software packages,
PAX is realized in the C++ programming language \cite{CPPSTL}.
It provides additional functionality in top of the vector algebra of the 
widely-spread libraries CLHEP \cite{CLHEP} or ROOT \cite{ROOT}.\footnote{
   At compile-time, the user can choose between the vector algebra packages of 
   CLHEP \cite{CLHEP} (default) or ROOT \cite{ROOT}.
   Depending on a compiler switch, the two type definitions \PaxLV\ and \PaxTV\
   are set to \CClass{HepLorentzVector} and \CClass{Hep3Vector} of CLHEP 
   or to \CClass{TLorentzVector} and \CClass{TVector3} of ROOT.}
The PAX container model as well as file I/O are based on the C++ 
Standard Template Library (STL) \cite{CPPSTL}. 

The PAX toolkit provides three types of generalized physics objects:
\begin{itemize}
\item{particles (or reconstructed objects), i.e.\ Lorentz-vectors, 
      represented by the class \PaxFV,}
\item{vertices, i.e.\ three-vectors,
      represented by the class \PaxVX,}
\item{and collisions, represented by the 
      class \PaxCO.}
\end{itemize}
These objects are able to establish relations, and can be stored 
and managed in event containers, represented by the \PaxEI\ class. 

\subsection{Physics objects}

\FigIns{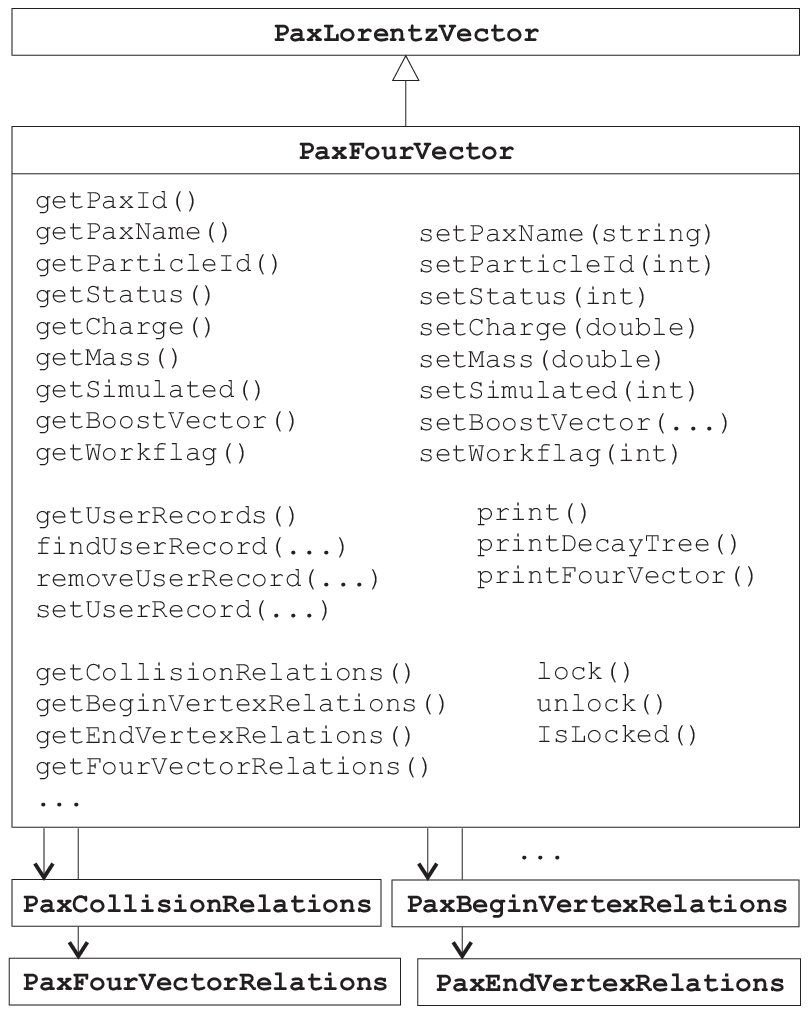}
       {The \PaxFV\ class extends the basic functionalities of the \PaxLV\ in order 
        to represent particles in HEP decays.}

\FigIns{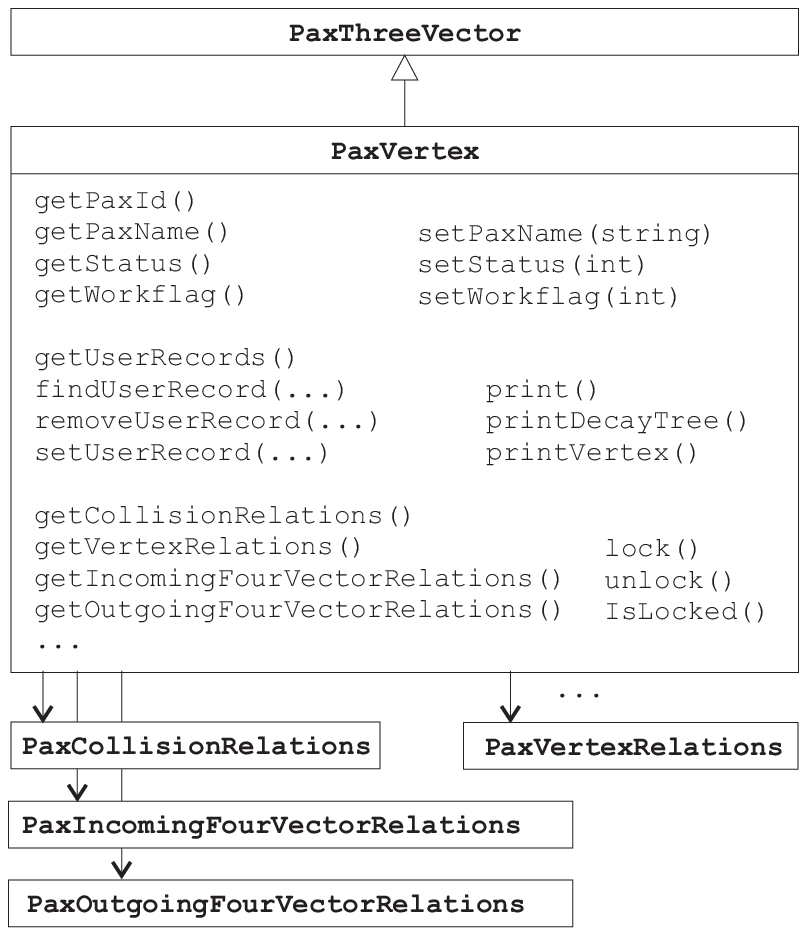}
       {The \PaxVX\ class extends the basic functionalities of the \PaxTV\ in order 
        to represent vertices in HEP particle decays.}

\FigIns{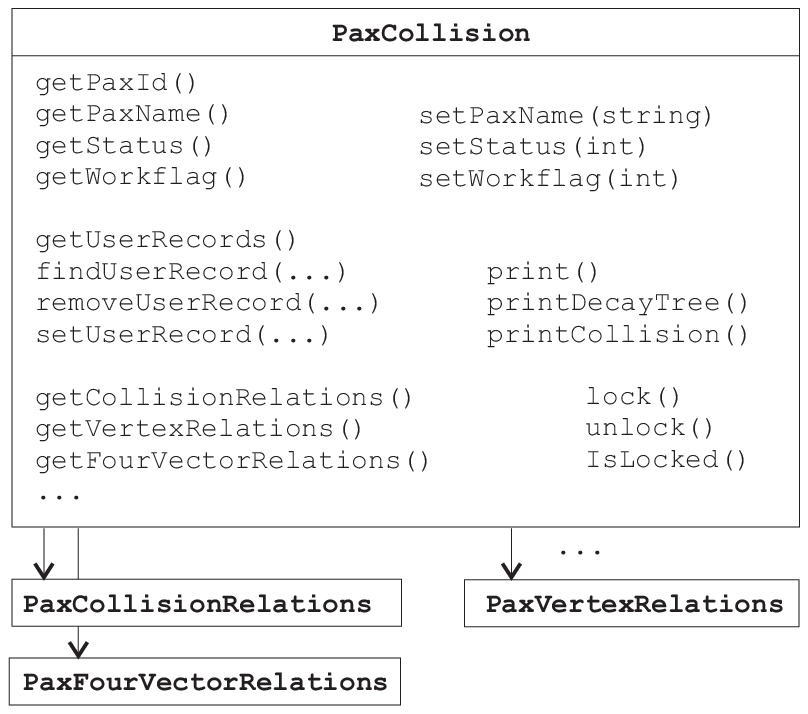}
       {The \PaxCO\ class represents collisions in bunch crossings at 
        high luminosity colliders.
        Besides storage of general properties, the \PaxCO\ allows the user to 
        establish and manage relations to \PaxVX\ and \PaxFV\ objects.}

The \PaxFV\ class (see \Fig{PaxFourVector.eps}) 
represents particles or reconstructed objects 
(such as muons, electrons, missing energy, jets etc.). 
It inherits its basic Lorentz-vector characteristics from the 
well-known libraries CLHEP or ROOT. Commonly needed, additional 
properties such as particle-id, status, charge etc.\ can be stored in
data members. Specific information (such as b-tags, 
jet cone sizes or energy corrections, for instance) can be stored
in the so-called user records. 
User records are collections of string-double pairs, meant to hold 
object information complementary to data members.

All PAX physics objects own user records (instances of the class \PaxUR) 
and provide methods for quick access to individual user record entries. 
Each instance of a PAX physics object carries an unique integer key 
(the so-called \PaxID) and a string name (the so-called \PaxName). 
An integer workflag facilitates tagging of individual objects.
Print methods are provided to allow monitoring of object state and 
established relations on various verbosity levels. 
Copy constructors are provided to perform deep copies of PAX
physics objects.

The \PaxVX\ class, sketched in \Fig{PaxVertex.eps}, represents 
the spatial point of decays in particle
reactions. Thus, in analogy with the \PaxFV, it obtains its basic 
three-vector characteristics also from the CLHEP or ROOT package.

The \PaxCO\ class  (see \Fig{PaxCollision.eps}) allows the 
separation of collisions in multicollision events, as they occur 
at high-rate hadron colliders.
It provides the relation management necessary to 
associate \PaxVX\ and \PaxFV\ objects with different collisions
in the event.

\subsection{Access to primordial C++ classes}\label{sectionExpClassRel}

Each PAX physics object can record pointers to an arbitrary 
number of instances of arbitrary C++ classes. This way, the 
user can keep track of the data origin within the detector 
reconstruction software, for instance. Access to the 
pointers is possible at the same runtime
during any later stage of the analysis. A typical use case is the 
need to re-fit a track which requires access to the hits in the 
tracking chamber. The PAX object that represents this track,
i.e.\ a \PaxFV\ instance, provides the two template methods
\CClass{addPointer$<$Type$>$(name, ID, pointer)}
  and 
\CClass{findPointer$<$Type$>$(name, ID)}.
The argument \CClass{name} is supposed to correspond to the 
C++ class name, e.g.\ \CClass{Type}, the argument \CClass{ID} is a unique integer 
identifier for the referenced instance of the C++ class \CClass{Type}, and the 
third argument is a pointer to this instance.

The mechanism behind is sketched in \Fig{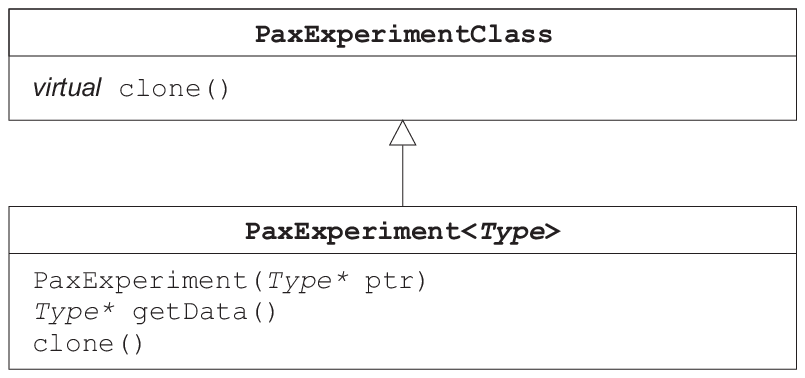}. 
The class template \CClass{PaxExperiment$<$Type$>$} provides storage, access, 
and clone of the pointer of type \CClass{Type}. Its base class 
\PaxEC\ is used as the interface to the PAX 
classes which are enabled to store and access the pointer through 
the C++ \verb+dynamic_cast+ operator. 

When copying a PAX physics object, all pointers are copied as well 
by making use of the \CClass{clone()} method.

\FigIns{PaxExperimentClass.eps}
       {The classes 
        \PaxEC\ and 
        \CClass{PaxExperiment$<$Type$>$}\
        provide recording of arbitrary pointers 
        with PAX objects.}

\subsection{Relation management}

The principal duty of the PAX relation management is
handling of decay trees. The manager is based on
the Mediator design pattern, described in detail in reference \cite{Mediator}.
In this design all relations are kept locally (i.e.\ every object knows 
about all their directly related objects), so that global 
relation directories can be avoided.

\FigIns{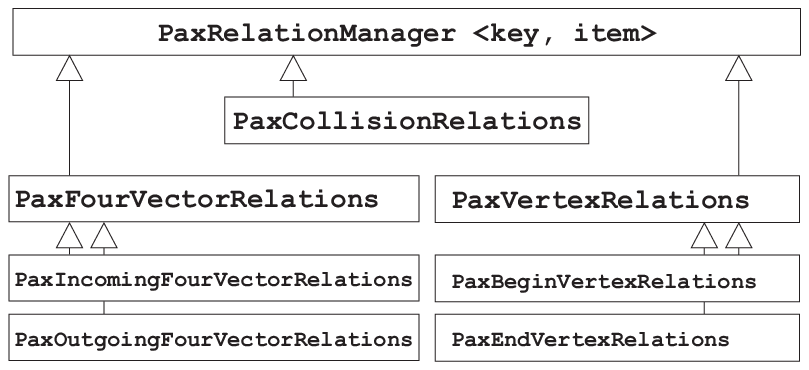}
       {The PAX classes for relation management inherit from the 
        class \PaxRelationManager.}

Speaking of PAX physics objects, this means, that each \PaxCO\ object 
owns relation managers (see \Fig{PaxRelMgr.eps}) that carry pointers to 
the related \PaxVX\ and \PaxFV\ objects. 
At the same time, the \PaxVX\ objects hold pointers to their related 
\PaxCO s as well as to their incoming and outgoing \PaxFV s. 
By the same token, \PaxFV s know about their related \PaxCO s and 
about their begin and end \PaxVX\ objects. With this functionality, 
PAX allows to store complete multicollision events from parton
to stable particle level, including four-momenta and spatial vertex 
information.

In addition, the PAX relation management is used to record
analysis histories: each object, which is copied via copy constructors,
keeps pointers to its original instances.
This way the user may always go back and ask for
original properties of objects which might have changed during
the development of the analysis.

A powerful feature, implemented by means of the relation management, 
is the so-called locking mechanism. It is implemented to enable the 
user to exclude parts of decay trees from the analysis (i.e.\ excluding
a lepton from a jet finding algorithm, etc.). If one particle 
or vertex is locked, all the objects down the decay 
tree (and the history) will be locked, too. 
Locking and unlocking are relaized by setting or removing the
lock-flag owned by each PAX physics object.

\subsection{Maps \& object containers}

The PAX kernel provides the base classes \PaxMapKI{key}{item}\
and \PaxMultiMapKI{key}{item},
which inherit from the STL classes \CClass{map$<$key, item$>$} and
\CClass{multimap$<$key, item$>$}, respectively. 
The explicit inheritance has been chosen to provide 
the use of existing STL objects and methods with these PAX 
classes. 
This way, iterations of PAX maps can be performed by using either 
the PAX iterator classes (\PaxIterator, \PaxMapIterator, \PaxMultiMapIterator) 
or the commonly known STL iterators. All PAX classes which
serve as containers are based on the class \PaxMap\ (see 
\Fig{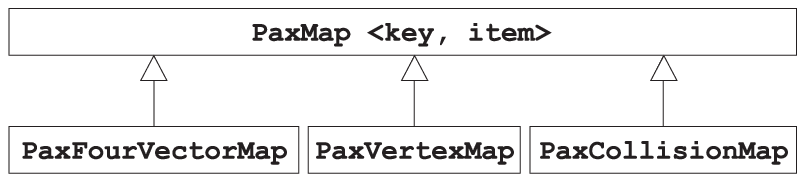}).

\FigIns{PaxContainers.eps}
       {The PAX container classes inherit from the class \PaxMap.}

\subsection{Event container}

\FigIns{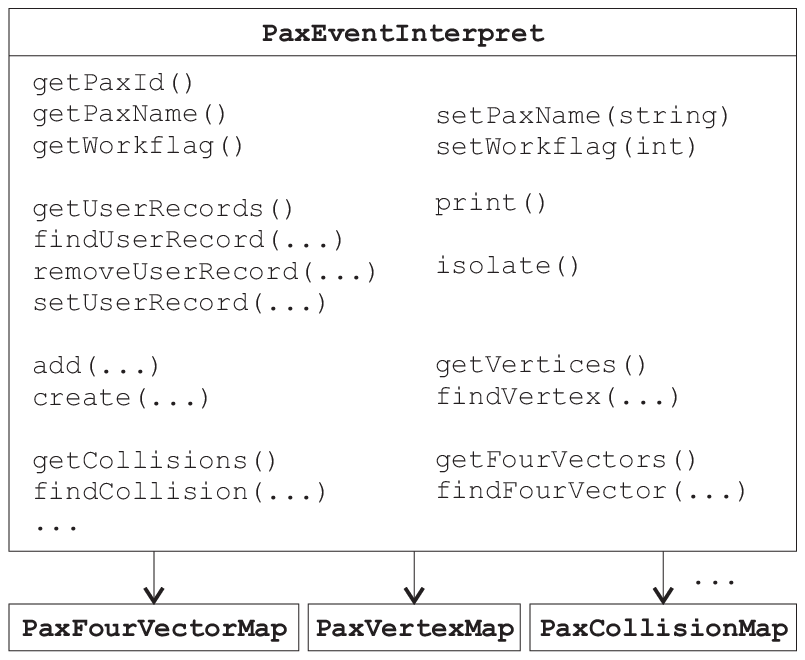}
       {The \PaxEI\ class represents the
        generalized container for complete HEP
        events. It stores and handles multiple collisions, vertices and particles
        as well as event specific information in the user records.}

The \PaxEI\ class, illustrated in \Fig{PaxEventInterpret.eps}, is the 
generalized event container provided by PAX. 
By incorporating the previously described functionalities, it is capable of 
holding the complete information of one multicollision event with
decay trees, spatial vertex information, four-momenta as well as 
additional reconstruction data in the user records. 
Physics objects (i.e.\ instances of the classes \PaxFV, \PaxVX\ and \PaxCO)
can be added or created with the
\PaxEI\CMethod{add()} and \PaxEI\CMethod{create()} methods.
Depending on the object type, a pair of \PaxID\ and Pointer to the 
individual object is stored in one of three maps (\PaxFVM, \PaxVXM\ or \PaxCOM).
Access to these maps as well as direct access to the physics 
objects is guaranteed via methods such as \PaxEI\CMethod{getFourVectors()}
and \PaxEI\CMethod{findFourVector()}. 
At deletion of a \PaxEI\ instance, all contained physics objects 
will be deleted, too.

The \PaxEI\ class is so named, because it is intended to
represent a distinct interpretation of an event configuration (e.g.\
connecting particles to the decay tree according to one out of a
number of hypotheses, applying different jet energy corrections, etc.). To facilitate the development of numerous
parallel or subsequent event interpretations, the \PaxEI\ class
features a copy constructor, which provides a deep 
copy of the event container with all data members, 
physics objects, and their (redirected) relations.

\subsection{PAX file I/O}\label{sectionPaxIoFile}

The PAX toolkit offers a file I/O scheme for persistent
storage of the event container, based on STL streams. 
It allows the user to write the contents of 
\PaxEI\ instances with all contained physics objects\footnote{
   For obvious reasons, pointers recorded with 
   PAX physics objects by means of the \PaxEC\ 
   functionality (as described in section \ref{sectionExpClassRel})
   are not stored to disk.
   } 
as well as their relations to PAX data files. When restoring
the data from file, an empty \PaxEI\ instance is filled with 
the stored data and objects and all object relations are 
reproduced.

The PAX data file format provides multi-version and 
multi-platform compatibility. It is built of a hierarchy 
of binary data chunks: the top level unit is an event, which 
consists of an arbitrary number of event interpretations. 
The event interpretation chunk consists of data members,
user records as well as chunks for each of the contained 
physics objects. Each chunk carries header 
information (one byte for unit type and four bytes for data 
amount information) and the actual binary data. 
This allows file structure checks and fast positioning. 
Therefore, the user can quickly skip arbitrary numbers of 
events in PAX data files, without having to sequentially 
read and discard. 

PAX also provides the possibility to write event units to 
strings (and to restore the \PaxEI\ instances from those 
strings). This way, the user can store PAX objects to any 
data format supporting strings or binary data fields (like 
databases or experiment specific data formats).

\subsection{Accessories and interfaces}

As a complement to the PAX kernel, we released two accessory 
packages for reading standard event generator file formats. 
The \CClass{PaxTuple} package provides transfilling of 
decay trees stored in the HEPEVT or ROOT Ntuple 
data formats to \PaxEI\ containers. 
Accordingly, the \CClass{PaxHepMC} package gives access to 
HepMC files. 

In addition, interfaces developed and posted by PAX users, 
that fill PAX objects with specific data of HEP experiments, 
are available via the PAX web page \cite{PAXWWW}.

\subsection{Software development procedure}

The PAX kernel and its officially supported accessories
are coded and maintained by a core group of currently 
six developers at CERN and the Aachen and Karlsruhe universities. 
New developments and code modifications pass a certification
procedure and are discussed and adopted in regular video meetings. 
As a guideline, new developments focus on aspects 
of performance improvement and on user feedback. 
New releases are to be backward compatible. Version 
management of the software project is handled with a
web-browsable Version Control System (CVS) \cite{CVS}\cite{PAXCVS}.

\subsection{Availability, documentation and support}

The continuously updated PAX web page \cite{PAXWWW} provides
download of the various versions of PAX kernel and accessories 
(based on the aforementioned web-browsable CVS repository).
It also provides the PAX Users Guide\cite{PAXGuide}, a comprehensive text 
documentation of the PAX toolkit, as well as class reference 
and fast navigator pages for download or online use.
The web page also offers access to mailing lists, in which PAX
users are informed about new developments and in which
technical issues of PAX analyses can be discussed.

\section{How PAX physics analyses can be structured}

\FigIns{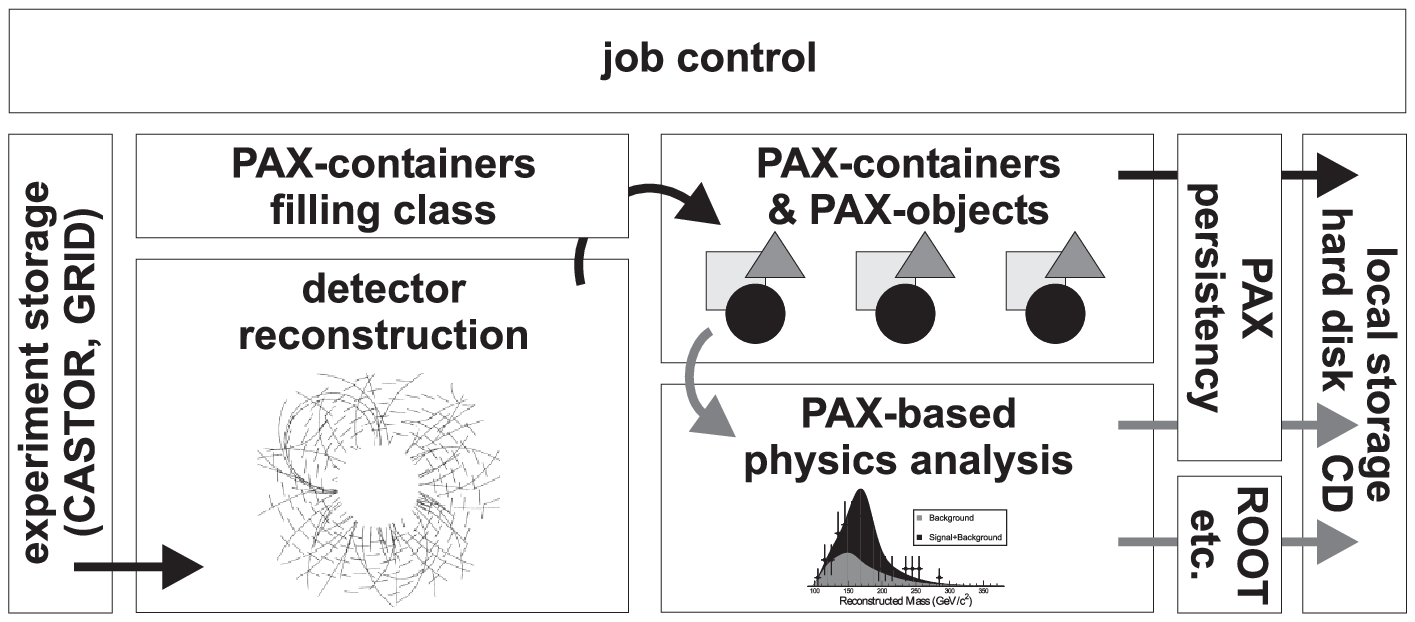}
       {One possible realization of a physics analysis with PAX; a dedicated,
        experiment-specific class for filling the PAX containers
        represents the interface between detector reconstruction software and PAX-based physics analysis.
        The PAX persistency scheme is used to store the data to PAX data files for later use.}
\FigIns{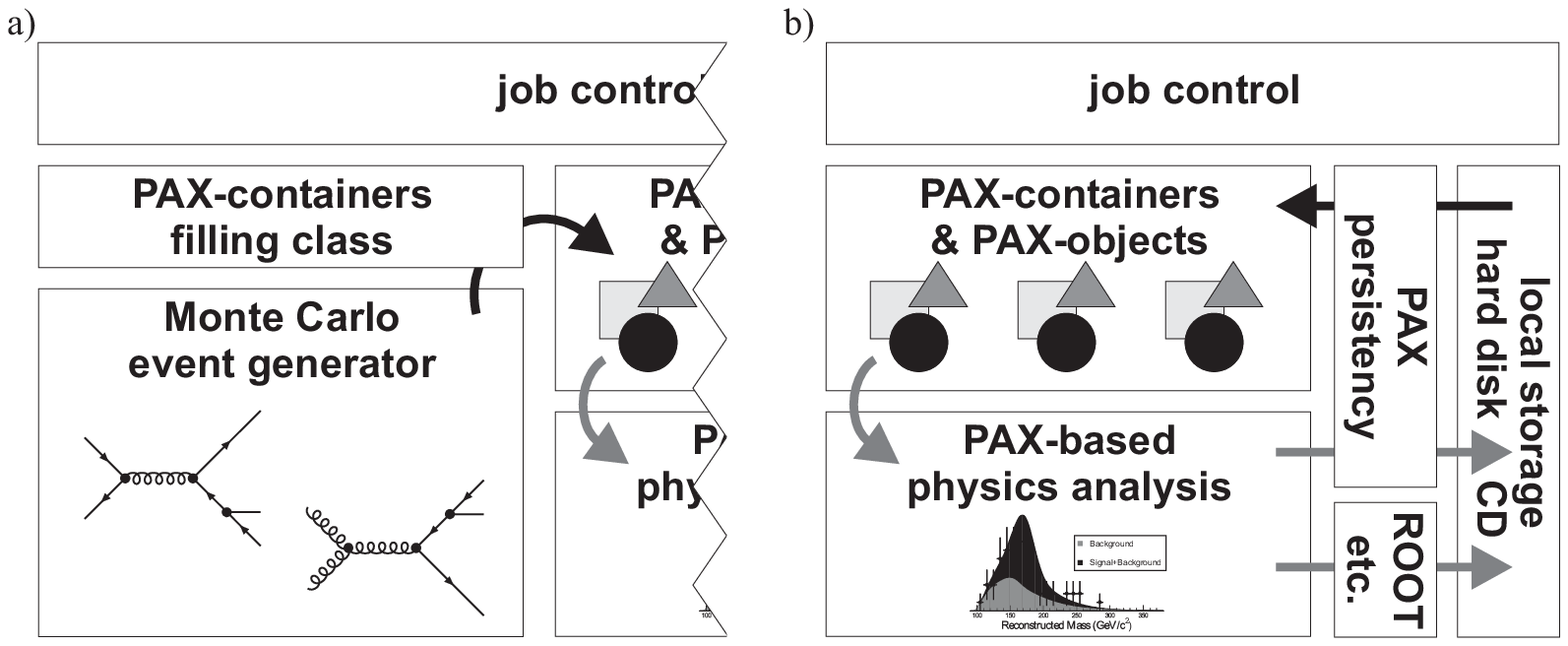}
       {a) Exchangeability of the filling class allows PAX physics analyses
           to be applied to various input sources, e.g.\ to Monte Carlo event generator data.
        b) The use of PAX data files allows fast analysis of the reconstruction
           data decoupled from the experiment-specific environment.}

To exploit the features offered by the PAX toolkit, 
physics analyses might be realized, for instance, according to 
the example structure illustrated in \Fig{PAXAnaI.eps}.

There, a dedicated, experiment-specific interface class for filling
the PAX containers (i.e.\ \PaxEI\ instances) represents the 
interface between detector reconstruction software and PAX-based 
physics analysis.
Once all relevant information is filled, the analysis code is called, 
and the PAX objects (as obtained by the filling class or at any subsequent
stage of the event analysis) can be stored persistently to PAX data 
files for later use. Analysis results might be further processed with 
help of the ROOT package.

With an analysis consistently formulated with PAX objects, the
filling class can be exchanged easily, and the identical analysis
code can be applied, for instance, directly to the output of a
Monte Carlo event generator or a fast simulation software,
see \Fig{PAXAnaII.eps}.a.
Furthermore, the use of PAX data files, which provide the distilled
experimental event information, allows fast analysis of the
reconstruction data decoupled from the experiment-specific software
and data storage environment, see \Fig{PAXAnaII.eps}.b.

\section{Implementation of PAX into experiment specific software environments}

PAX has been made available within the software environments of the 
experiments CDF, D0\footnote{Interfaces to the D0 software 
   are available as $\beta$-version since April 2005.} 
(both Tevatron) and CMS (LHC). 

Following the same principles, the integration of PAX into the latter is 
described as a general example.

The PAX toolkit is provided by the CMS software environment
as an external package \cite{PAXAFS}, enabling the physicists
inside the CMS collaboration to use PAX without having to care
about installation or setup of the package.

An extensive example analysis for the use of PAX with the detector
reconstruction software ORCA \cite{ORCA} is included in the
CMS CVS repository \cite{PAXExPaxAnalysis}. In this example,
the (ambiguous) reconstruction of the partonic process of the decay
$W\decaysto \mu \bar \nu_\mu$ is carried out by using reconstructed
muons and missing transverse energy. The missing information about
the longitudinal component of the neutrino momentum is obtained with
a $W$-mass constraint, which yields (up to) two analytical solutions,
and thus two possible event interpretations. Subsequently, both
interpretations are developed in two separate \PaxEI\ instances,
and a number of example histograms is filled.

The class design of this example analysis is based on the structure
described in the previous section, including interface classes
for filling \PaxEI\ containers with the reconstructed objects
of ORCA.

To facilitate the start-up for new PAX users, a tutorial video for this
example plus supplementary material can be found in the CMS section of
the PAX web page \cite{PAXTutorial}.

\section{PAX physics analyses for Tevatron and LHC}

Provided for the software environments of the CDF, D0 and CMS 
experiments, PAX is being explored by a growing user community. 
In the following, two successful applications of PAX in
complex physics analyses are presented.

\subsection{A PAX-based $t \bar t$ analysis for CDF}

\FigIns{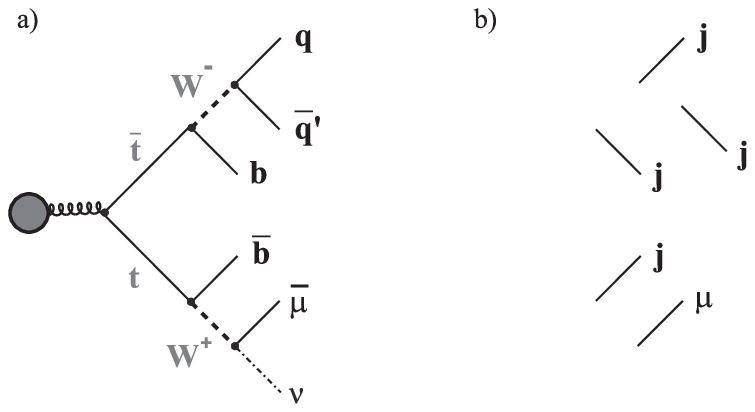}{The channel $t \bar t$ on parton level (a) and the visible reconstructed partons of this channel (b).}
\FigIns{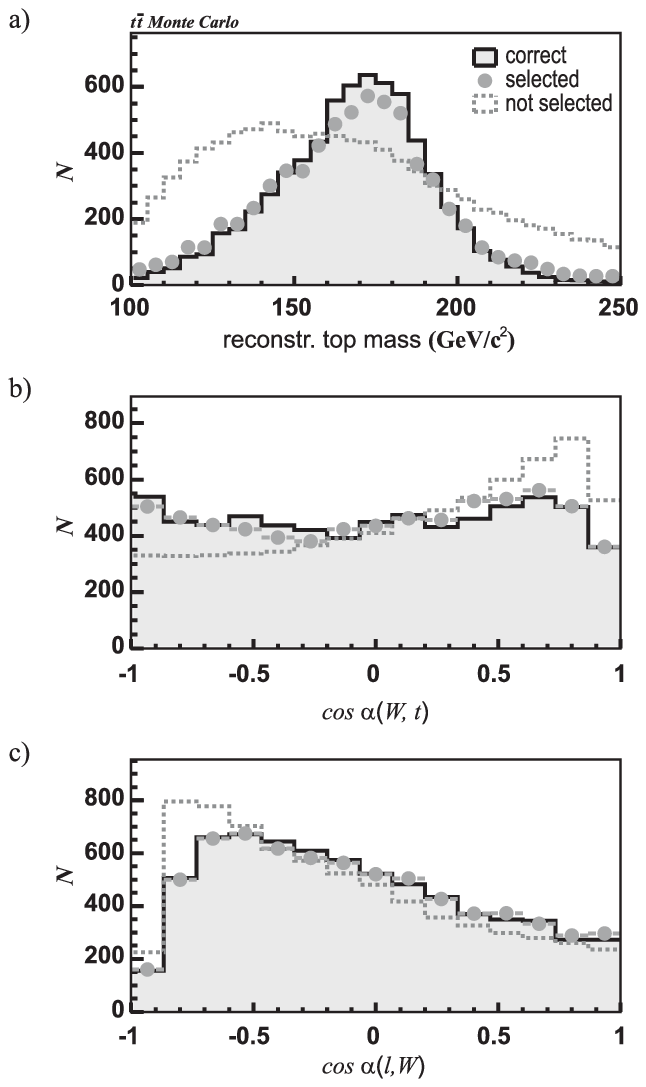}{
           Verification of the $t$-quark reconstruction in generated $t \bar t$ events.
           The full histograms show reconstructed properties of the event interpretation 
           which reproduces the partonic $t \bar t$ state best. Further information 
           results from the selection procedure using reconstructed quantites of the 
           event only: the symbols represent the selected event interpretation, the 
           dashed histogram summarizes the other possible interpretations.
           a) Reconstructed mass of the $t$-quark with a subsequent leptonic $W$-decay.
           b) Angular distribution of the $W$-boson in the rest frame of the $t$-quark.
           c) Angular distribution of the charged lepton in the rest frame of the $W$-boson. 
           (For this study, the HERWIG Monte Carlo generator \cite{HERWIG} and
            CDF detector simulation \cite{CDFMC} have been used.)}

In this section, an analysis of top-antitop-quark 
events ($t \bar t$ events) with the CDF experiment at Tevatron is described
\cite{DHDiss}. As illustrated in \Fig{ttFeyn.eps},
the electron-plus-jet decay channel shows similar
combinatorial tasks as the aforementioned $\ttH$
channel.

In this $t \bar t$ study, an analysis factory based on
the PAX event interpretation concept is used
to perform complete reconstruction of the
partonic scattering process and to optimize
the separation of signal and background
processes. 

The partonic process of the decay
$\bar t \decaysto W \bar b \decaysto e \bar \nu_e \bar b$
is reconstructed as follows.
First, the W-boson decaying into electron and neutrino
is reconstructed. From the W-mass constraint two
possible solutions can be deduced for the longitudinal
neutrino momentum. This results in two event
interpretations for the W-boson. Combining each of those
with one of the jets leads to the interpretations
for the $t$-quark (with different kinematics and reconstructed masses).
The remaining part of the process, i.e.\ $t \decaysto Wb \decaysto q\bar q'b$,
is reconstructed from three of the remaining jets.
Consequently, in a four jet $t\bar t$ event, 24 interpretations can
be constructed.
The most likely $t\bar t$ event interpretation is selected by
first demanding non-zero b-probability
for one of the jets of one of the $t$-quark candidates.
Finally, one of these solutions is selected by 
evaluating the most likely event interpretation based 
on kinematic properties, the reconstructed mass of the 
W boson decaying to $q\bar q'$, and the mass difference of 
the two reconstructed $t$-quarks.
The resulting example plots are shown in \Fig{ttResults.eps}.

\subsection{A PAX-based $\ttH$ analysis for CMS}

\FigIns{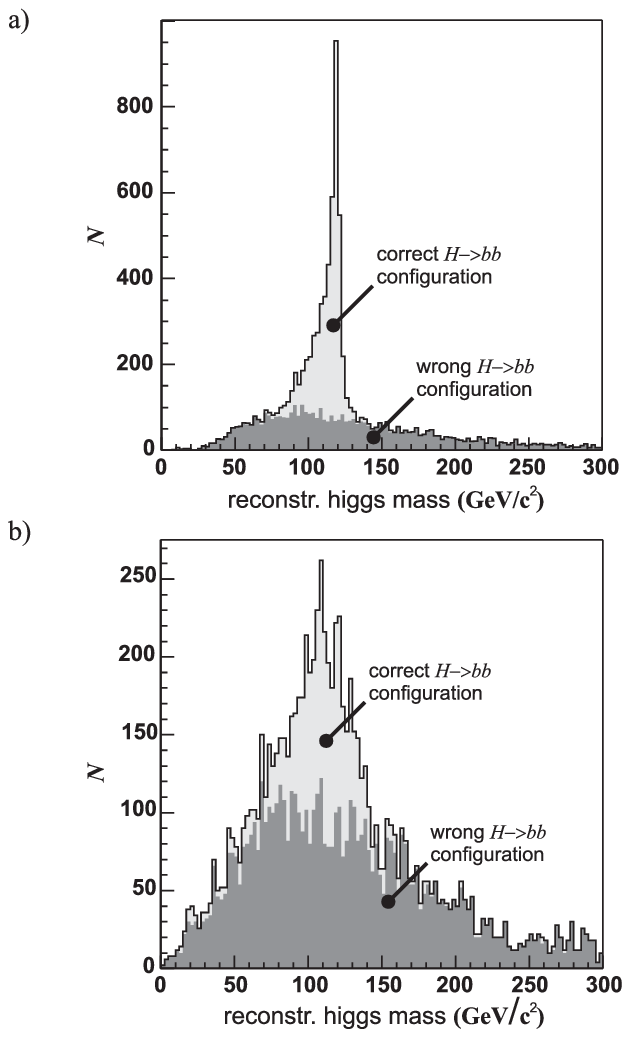}{Reconstructed Higgs mass in the channel $\ttH$ with $\Hbb$
                        on generator (a) and full simulation level (b).
                        The gray shaded area corresponds to the combinatorial background, 
                        i.e.\ to those events, in which a wrong $\Hbb$ configuration was 
                        selected.
                       (For this study, the PYTHIA Monte Carlo generator \cite{PYTHIA} and
                        CMS detector simulation \cite{CMSMC} have been used.)}

The channel of associated Higgs production, $\ttH$ with $\Hbb$, by means
of which the requirements to a particle physics analysis
toolkit have been motivated in the introduction of this article,
is studied in the CMS experiment at the LHC \cite{HiggsDiscovPot}\cite{SKDiss}, 
for instance.

The most recent of these studies makes use of the PAX event 
interpretation concept to develop possible event 
interpretations in a manner similar to the one described in the 
previous CDF example. After development of all interpretations, 
a likelihood function is used to select the 
most probable one by rating the different configurations on the basis 
of kinematics variables and masses of the two $t$-quarks and their 
decay products.

\Fig{ttHResults.eps} illustrates the performance of this method in simulations
with and without detector effects. Please notice, that \Fig{ttHResults.eps}.a 
and \Fig{ttHResults.eps}.b  have been produced with the identical analysis code, 
by simply exchanging the interface classes (compare \Fig{PAXAnaI.eps} and \Fig{PAXAnaII.eps}). 
In this way, a good measure for how detector and reconstruction methods influence the results can directly be 
obtained -- with almost no analysis code duplication.

\section{Conclusions}

The PAX toolkit is designed to assist physicists at modern collider experiments
in the analysis of complex scattering processes. PAX provides a generalized 
HEP event container with three types of physics objects (particles, 
vertices and collisions), relation management and file I/O scheme. 

The PAX event container is capable of storing the complete information of 
multicollision events (including decay trees with spatial vertex information, 
four-momenta as well as additional reconstruction data). 
An automated copy functionality for the event container allows the user to 
consistently duplicate event containers with physics objects and relations. 
The PAX file I/O scheme can be used to write (and read) complete event containers 
to (from) disk file; this offers an easy realization of distilled experiment 
data streams.
By structuring physics analyses based on PAX objects, 
the identical source code can be applied to various data 
levels. This adds a desirable aspect of flexibility to the 
software-side of particle physics analysis.

PAX is available within the software environments of experiments 
at Tevatron and LHC, where it is applied in a number of physics 
analyses. Two thereof are outlined in this article, demonstrating
typical use cases and successful applications of the PAX toolkit. 
Evident advantages arising from the usage of the PAX toolkit 
are avoidance of code duplication, increased code lucidity, 
unified data model and nomenclature, and therefore more efficient 
team work in the complex physics analyses at modern HEP experiments.

\section*{Acknowledgment}

The authors would like to thank 
Rene Brun,
Anne-Sylvie Giolo-Nicollerat, 
Christopher Jung, 
Yves Kemp, 
Klaus Rabbertz, 
Jens Rehn, 
Sven Schalla, 
Patrick Schemitz, 
Thorsten Walter, 
and Christian Weiser 
for helpful contributions and feedback.




\end{document}